\documentclass[12pt]{article}
\usepackage{graphicx}

\newcommand\pubnumber{Cavendish-HEP-22/10}
\newcommand\pubdate{\today}

\def\institute{Cavendish Laboratory Cambridge University\\
JJ Thomson Ave, Cambridge CB3 0HE, UK}
\def\support{\footnote{Leverhulme Trust and the Isaac Newton Trust.}}

\def\Title#1{\begin{center} {\Large #1 } \end{center}}
\def\Author#1{\begin{center}{ \sc #1} \end{center}}
\def\Address#1{\begin{center}{ \it #1} \end{center}}

\newcommand\pubblock{\rightline{\begin{tabular}{l} \pubnumber\\
         \pubdate  \end{tabular}}}
\newenvironment{Abstract}{\begin{quotation}  }{\end{quotation}}
\newenvironment{Presented}{\begin{quotation} \begin{center} 
             PRESENTED AT\end{center}\bigskip 
      \begin{center}\begin{large}}{\end{large}\end{center} \end{quotation}}
\def\Acknowledgements{\bigskip  \bigskip \begin{center} \begin{large}
             \bf ACKNOWLEDGEMENTS \end{large}\end{center}}

\def\beq{\begin{equation}}
\def\eeq#1{\label{#1}\end{equation}}
\def\eeqn{\end{equation}}

\def\beqa{\begin{eqnarray}}
\def\eeqa#1{\label{#1}\end{eqnarray}}
\def\eeqan{\end{eqnarray}}

\let\bar=\overbar

\def\Dslash{\not{\hbox{\kern-4pt $D$}}}
\def\dslash{\not{\hbox{\kern-2pt $\del$}}}

\def\msb{{\bar{\ssstyle M \kern -1pt S}}}

\begin{document}
\begin{titlepage}
\pubblock

\vfill
\Title{Precision comparisons between theory and data in $t\bar{t}$-production at the LHC}
\vfill
\Author{Rene Poncelet\support}
\Address{\institute}
\vfill
\begin{Abstract}
NNLO QCD corrections to the production and decay of top-quark pairs allow performing precision phenomenology in inclusive and fiducial phase spaces.
State-of-the-art predictions for top-quark pair production and their comparison to recent collider data are discussed.
\end{Abstract}
\vfill
\begin{Presented}
$15^\mathrm{th}$ International Workshop on Top Quark Physics\\
Durham, UK, 4--9 September, 2022
\end{Presented}
\vfill
\end{titlepage}
\def\thefootnote{\fnsymbol{footnote}}
\setcounter{footnote}{0}

\section{Introduction}

The measurement of top-quark pair production cross-sections is a staple of Standard Model (SM) measurements at the Large Hadron Collider (LHC). The large number of top-quark pairs produced at the LHC allows for percent-level precision in determining total and differential cross-sections. These measurements can constrain SM parameters such as the top-quark mass, the strong coupling constant or parton distribution functions (PDF). Precise and accurate theory calculations, including next-to-leading (NLO) and even next-to-next-to-leading order (NNLO) corrections, are essential for comparing theory to data.

NNLO calculations for on-shell top-quarks \cite{Czakon:2015owf} are state-of-the-art in QCD. Recently, computations for on-shell and stable top-quarks have been extended to include the decay within the Narrow-Width-Approximation (NWA) with full spin-correlations up to NNLO QCD \cite{Behring:2019iiv, Czakon:2020qbd}. This has been further extended to incorporate the fragmentation of final-state partons into B-hadrons, allowing for a realistic final-state model in ref.~\cite{Czakon:2021ohs, Czakon:2022pyz}. Also, NNLO QCD calculations for top-quark pair production matched to parton-showers (PS) \cite{Mazzitelli:2020jio, Mazzitelli:2021mmm} within the MiNNLOPS prescription became available. For these, the top-quark decays are handled by the parton-shower. For off-shell top-quarks, NLO QCD and EW computations, including the matching to PS, are routinely discussed in the literature, see, for example, ref.~\cite{Bevilacqua:2010qb} and ref.~\cite{Denner:2016jyo}.

This contribution discusses the NNLO QCD corrections to final-state observables for top-quark pair production in the di-leptonic decay channel. The presented observables are an extension of the work published in ref.~\cite{Czakon:2020qbd} and computed for a phase space mimicking the experimental fiducial phase space in a recent CMS measurement \cite{CMS:2022uae}. The same article compares the predictions discussed here and the actual measurement.

\section{NNLO QCD predictions for leptonic observables}

We consider the process $pp \to t\bar{t} + X \to \bar{\ell} \nu b \ell \bar{\nu} \bar{b} +X$ in the NWA approximation. All spin-correlations are kept, and NNLO corrections to the production and the decays are included. For implementation details, refer to ref.~\cite{Czakon:2020qbd}. The top-quark mass is set to $m_t = 172.5$ GeV, and we chose $H_T/4$ for the renormalisation and factorisation scale. Following the fiducial phase space definitions in ref.~\cite{CMS:2022uae}, we apply the following selection criteria:
\begin{itemize}
    \item Leptons: $p_T(\ell) \geq 20$ GeV, $|y(\ell)| \leq 2.4$ and $m(\ell\bar{\ell})\geq 20$ GeV.
    \item Jets: anti-$k_T$, $R = 0.4$, with $p_T(j)\geq 30$ GeV and $|y(j)| \leq 2.4$.
    \item At least two b-tagged jets pass the above criteria. 
\end{itemize}
We compute the complete leptonic and b-jet observables presented in ref.~\cite{CMS:2022uae}, including reconstructed top-quark differential distributions. All results are available in electronic format\footnote{https://www.precision.hep.phy.cam.ac.uk/results/ttbar-decay/}.

\begin{figure}
    \centering
    \includegraphics[width=0.5\textwidth]{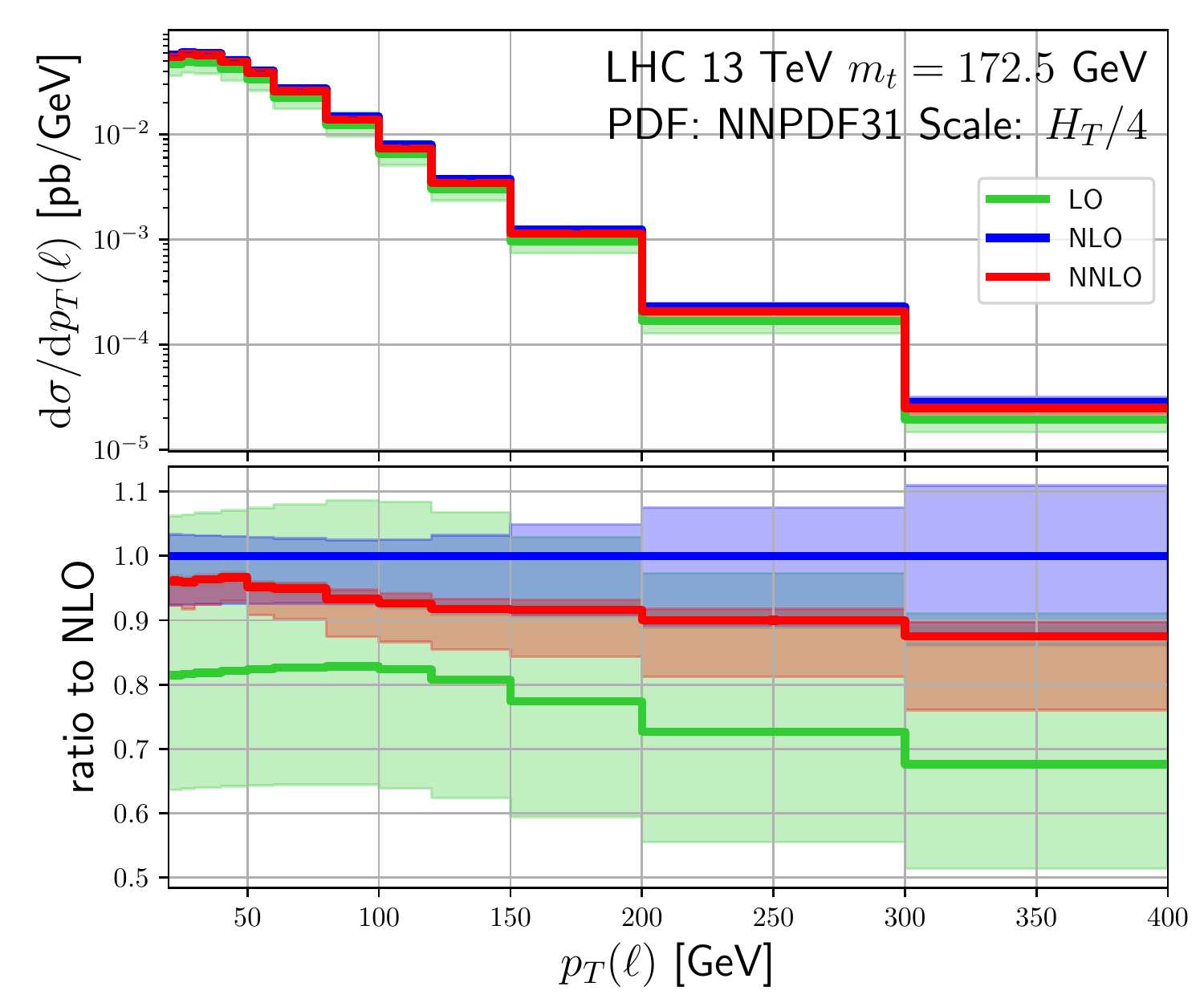}%
    \includegraphics[width=0.5\textwidth]{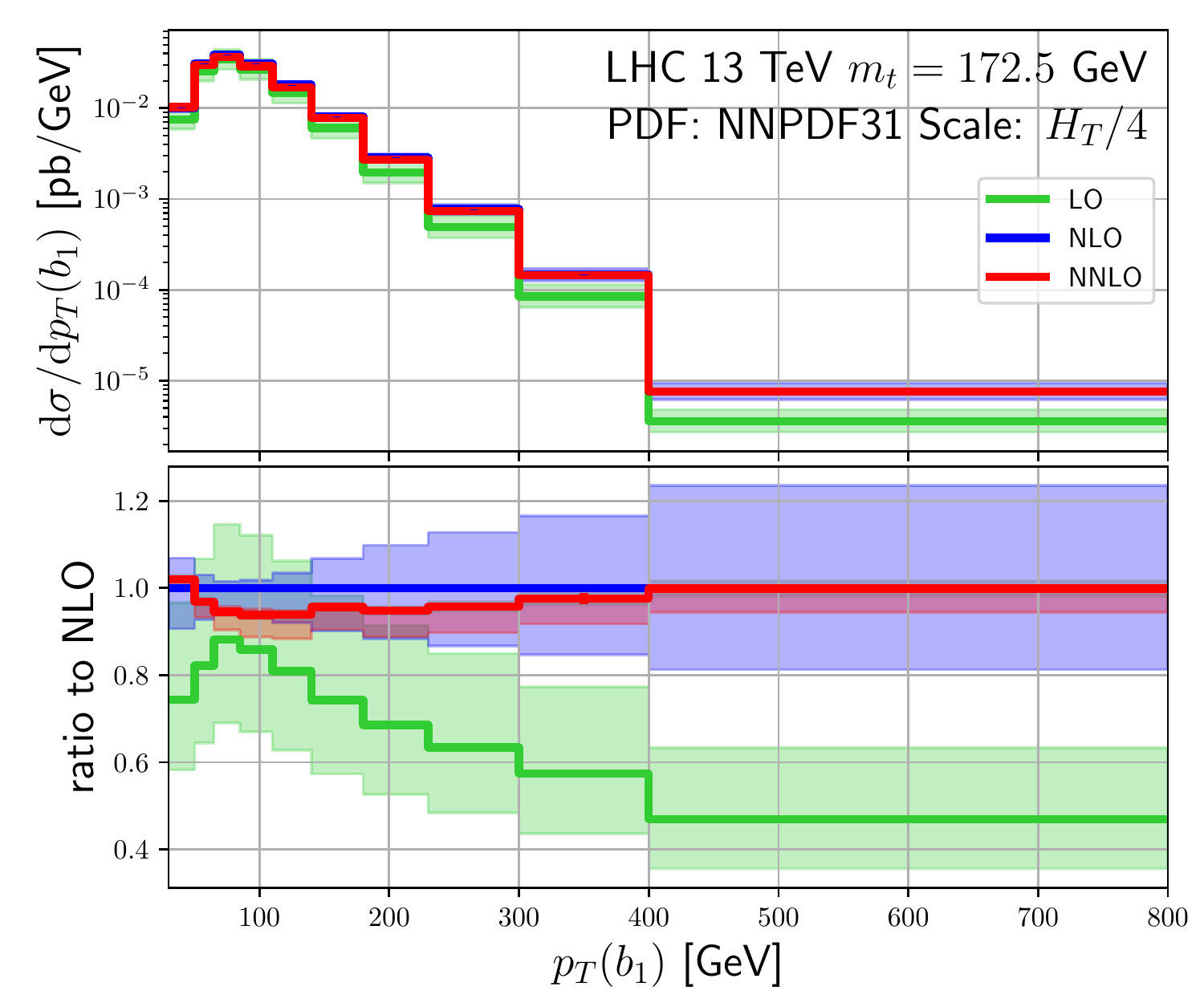}
    \caption{The transverse momentum of the charged lepton and the leading b-jet. Further description in the text.}
    \label{fig:fid1}
\end{figure}

As an illustrating example, Fig.~\ref{fig:fid1} shows the transverse momentum distribution of the charged lepton on the left and of the leading b-jet on the right. We show LO in light green, NLO in blue and NNLO QCD in red; the bands correspond to seven-point variation around the central renormalisation and factorisation scale. In the case of the charged lepton, we find negative NNLO QCD corrections ranging from $-5\%$ for low to $-10\%$ for high transverse momentum. The scale dependence decreases by a factor of 2 for low transverse momenta when going from NLO to NNLO QCD. At high transverse momenta, the reduction is smaller. For the leading b-jet transverse momentum, we find negative corrections of similar size around the peak of the distributions, while in the tail, the corrections become smaller. The scale dependence reduces significantly from about 20\% at NLO to 3-5\% at NNLO QCD in the high transverse momentum region. The comparison to data in ref.~\cite{CMS:2022uae} finds good agreement between these predictions and the 'particle-level' measurement.

\section{IR-safe flavoured anti-$k_T$ algorithms}

A subtle question concerning the NNLO QCD computation presented in ref.~\cite{Czakon:2020qbd} is the definition of the b-jets. The calculation implements a scheme with $n_l = 5$ light quarks, which implies  massless b-quarks. Due to the fixed-order nature of the computation, b-tags for jets are defined by the jet's parton content after clustering, essentially counting the number of (anti-)b-quarks and assigning a b-tag if the net bottomness is non-zero. Starting at NNLO QCD, this procedure is IR-unsafe \cite{Banfi:2006hf} due to configurations originating from soft gluons splitting to (massless) bottom-quark pairs. The computation in ref.~\cite{Czakon:2020qbd} employs a pragmatic solution by removing the singularities by a technical cut-off to avoid the numerical divergence associated with these configurations. Based on the fact that the numerical contribution from those configurations is small when computed with the help of massive b-quarks, this approximation is well-motivated, and the calculated cross-sections are not sensitive to the unphysical cut-off.

A more general solution to this issue is modifying the clustering algorithm to capture the IR-unsafe configurations. The flavoured $k_T$ algorithm presented in ref.~\cite{Banfi:2006hf} does have the necessary properties to define flavoured jets beyond NLO QCD. Unfortunately, it is impossible to directly translate the modifications to the anti-$k_T$ algorithm, the more commonly used prescription at the LHC. Recently, this topic found renewed interest leading to a range of proposed flavoured anti-$k_T$ prescriptions \cite{Caletti:2022hnc, Czakon:2022wam, Gauld:2022lem}.

\begin{figure}
    \centering
    \includegraphics[width=0.5\textwidth]{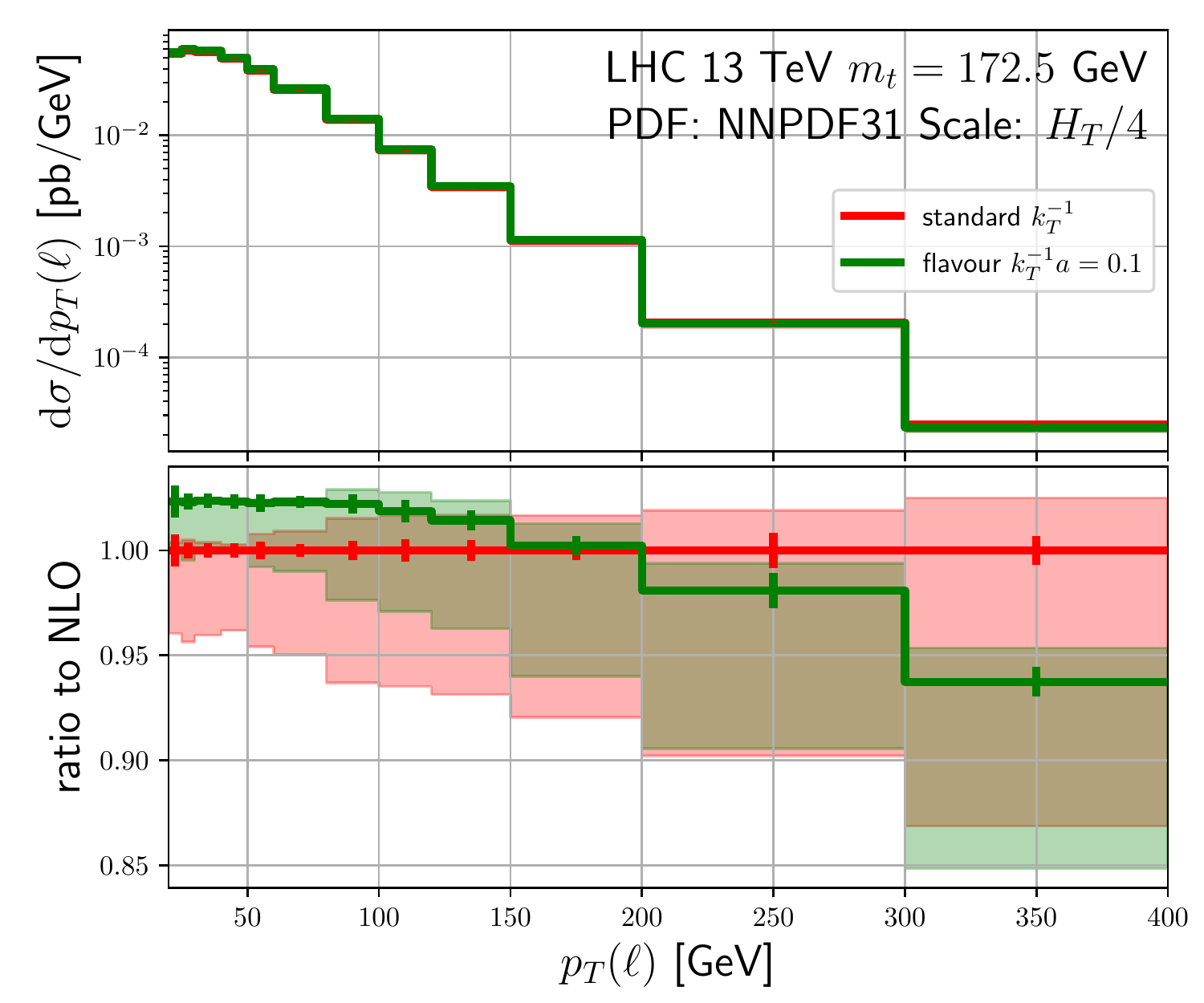}%
    \includegraphics[width=0.5\textwidth]{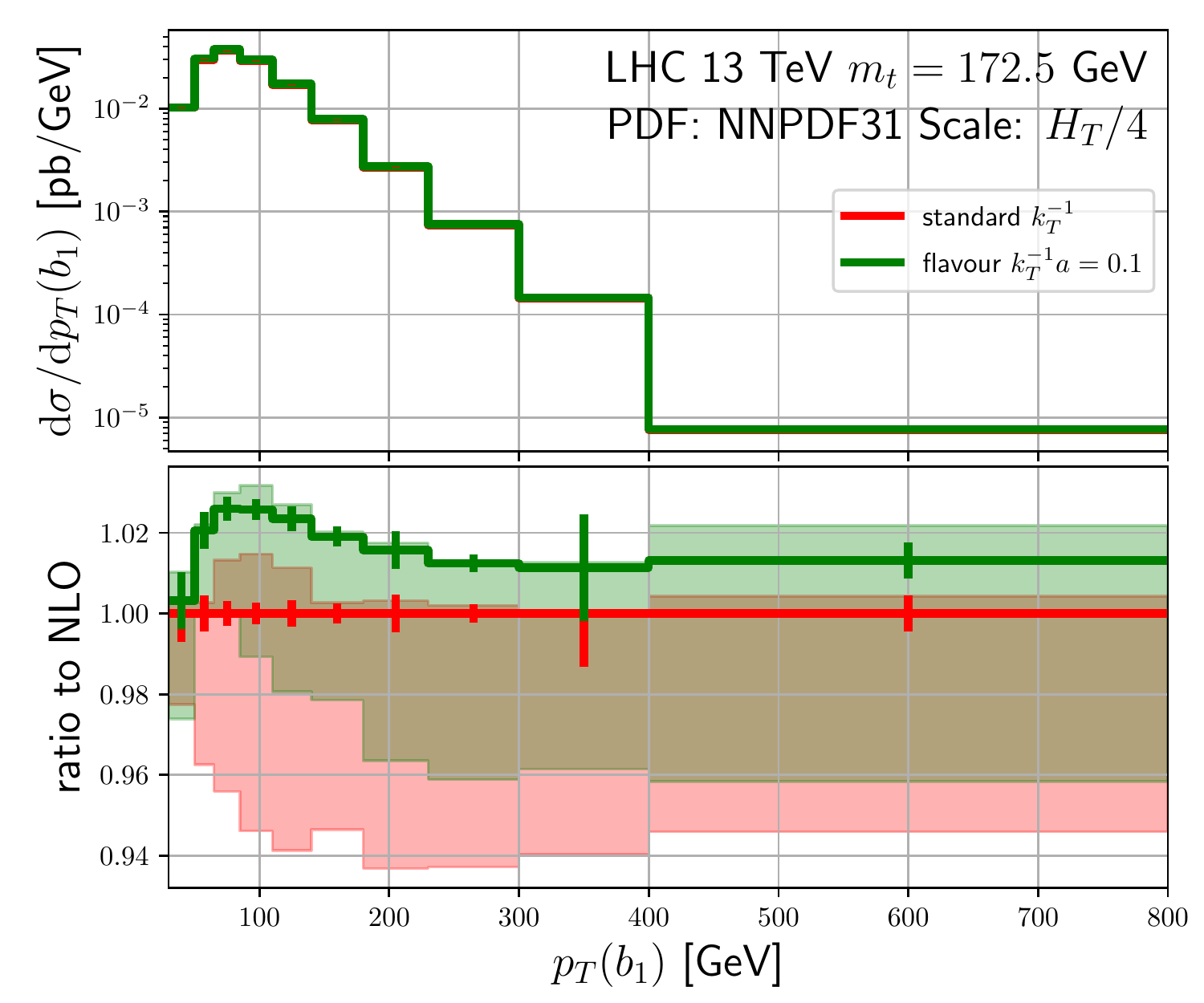}
    \caption{Comparison of anti-$k_T$ and flavoured anti-$k_T$ predictions at NNLO QCD for the transverse momentum of the charged lepton and the leading b-jet. Further description in the text.}
    \label{fig:fid2}
\end{figure}

Ref.~\cite{Czakon:2022wam} shows the application of the flavoured anti-$k_T$ algorithm to the process of top-quark pair production and decay and compares it to the cut-off prescription employed in ref.~\cite{Czakon:2020qbd}. Figure \ref{fig:fid2} shows the transverse momentum spectrum of the charged lepton (left) and the leading b-jet (right). The observed differences are minor and well within the remaining scale uncertainties, indicating that the IR-unsafe configurations' contribution is negligible in this case.

\section{Conclusion and outlook}

We computed new NNLO QCD predictions for top-quark pair production and decay in fiducial phase spaces, an extension of the work presented in ref.~\cite{Czakon:2020qbd}. The comparison to recent LHC data has been performed in ref.~\cite{CMS:2022uae}. Further, the effect of applying new flavoured jet algorithms has been investigated and found negligible for this process.

Extrapolation to the inclusive stable on-shell top-quark phase space is an important source of uncertainty in top-quark physics. Still, it is conceptually avoidable by studying differential distributions of the decay products within fiducial phase spaces. Higher-order predictions for such distributions are necessary to do so with precision. Working in fiducial phase spaces reveals old and new challenges in object definitions and unfolding.

\Acknowledgements
R.P. acknowledges the support from the Leverhulme Trust and the Isaac Newton Trust and the use of the DiRAC Cumulus HPC facility under Grant No. PPSP226.

\end{document}